\documentclass[10pt,twocolumn]{article}
\usepackage{graphicx}
\usepackage{amsfonts}
\usepackage{amsmath}
\usepackage{amssymb}
\usepackage{mathrsfs} 
\usepackage{lipsum}
\usepackage{color}
\usepackage{siunitx}
\usepackage{authblk}
\usepackage[margin=2cm]{geometry}
\usepackage{mathtools, cuted}
\usepackage[T1]{fontenc}
\usepackage{graphicx}
\usepackage{hyperref}
\usepackage{bm}

\begin{document}

\vspace{10pt}

\title{Blowing Big Bubbles}

\author{Christopher A.E. Hamlett,\textit{$^a$} Dolachai N. Boniface,\textit{$^b$} Anniina Salonen,\textit{$^b$} Emmanuelle Rio,\textit{$^b$} Connor Perkins,\textit{$^a$} Alastair Clark,\textit{$^a$} Sang Nyugen,\textit{$^a$} and David J. Fairhurst$^\ast$\textit{$^a$}\\
\small{\textit{$^{a}$~Nottingham Trent University, Nottingham, NG11 8NS, UK\\}
\textit{$^{b}$~Universit\'e Paris Saclay, CNRS, Laboratoire de Physique des Solides, 91405 Orsay, France.}}}
\date{\today}

\twocolumn[
   \begin{@twocolumnfalse}
        \maketitle
           \begin{abstract}
Although street artists have the know-how to blow bubbles over one meter in length, the bubble width is typically determined by the size of the hoop, or wand they use. In this article we explore a regime in which, by blowing gently, we generate bubbles with radius up to ten times larger than the wand. 
We observe the big bubbles at lowest air speeds, analogous to the dripping mode observed in droplet formation.
We also explore the impact of the surfactant chosen to stabilize the bubbles. We are able to create bubbles of comparable size using either Fairy liquid, a commercially available detergent often used by street artists, or sodium dodecyl sulfate (SDS) solutions. The bubbles obtained from Fairy liquid detach from the wand and are stable for several seconds, however those from SDS tend to burst just before detachment.
           \end{abstract}

   \end{@twocolumnfalse}]

To blow a soap bubble sparks joy in people of all ages, irrespective of their scientific knowledge. There is no need to understand Empress Dido's isoperimetric inequality \cite{Dido} to appreciate the simplicity of their perfectly spherical shape, arising from the minimisation of  surface energy. Nor is there a requirement to have read Newton's \textit{Opticks}\cite{newtonopticks} to be enchanted by the swirling bands of colours indicating how the thickness of the soap film  changes with drainage and evaporation \cite{isenberg1978science}. Street artists know that making stable giant bubbles is not easy: preparing the ideal soap solution for given environmental conditions (humidity, temperature, air speed) requires years of know-how. And although recent work \cite{frazier2020make} has highlighted how the addition of high-molecular weight polydisperse polymers can improve stability, there is no fundamental model  combining film formation and growth that can predict the optimal surfactant solution.
The analysis is further complicated when film rupture is also considered, a singular event \cite{rio2014thermodynamic} that is nonetheless essential for the bubble to detach from the original film: rupture controls when the blown film pinches off to produce a detached closed bubble\cite{boys1959soap,burton2005scaling}. 
To successfully blow individual bubbles they must not rupture until they have left the wand to drift away. 

The simple question of what determines the size at which the bubble detaches was first addressed by Plateau in the late 19$^{th}$ century \cite{plateau1873statique} and 80 years later Boys \cite{boys1959soap} presented a beautiful overview of early experimental measurements. However, it is only much more recently that Salkin \textit{et al.} \cite{salkin2016generating} designed a controlled experiment in which a bubble is formed by blowing a jet of air with controlled flow-rate through a film of falling surfactant solution.
At low air speeds the bubbles are generated one by one, while at higher air speeds the bubble size and frequency are set by the Rayleigh-Plateau instability \cite{salkin2014creation}.
Their results are applicable to both "contained" air jets, which fall entirely within the soap film, and "uncontained" jets which are larger than the film.
They show that the minimum velocity $v_0$ for which a bubble can be inflated is given by a simple expression found by equating the inertial pressure due to the jet of gas (density $\rho_g$) exiting from a nozzle of radius $R_0$  with the Laplace pressure 
exerted by the expanded air flow at a distance $d$ from the nozzle:
\begin{equation}
    v_0=\sqrt{\frac{8\gamma}{\rho_g R_0}\left(1+\frac{1}{5}\frac{d}{R_0}\right)}.
    \label{eq:V_0Salkin}
\end{equation} Turbulent jets, such as these, have a universal opening angle \cite{pope2000turbulent,labus1972experimental} of 11.8$^\circ \approx 1/5$ radians, so the jet diameter increases linearly with distance between nozzle and film.
In this regime, the bubble size is set by the Rayleigh-Plateau instability and equals $2R_0$. Su \textit{et al.} systematically varied the nozzle diameter and also found that bubble size was correctly predicted by Rayleigh-Plateau instability \cite{su2017investigation}. To generate large bubbles in the Rayleigh-Plateau regime, it is thus necessary to use a correspondingly large wand or a wide air flow.

In this article, we explore the possibility of making "giant" bubbles, significantly larger than the wand, by blowing downwards in the low air-speed regime. This follows on from preliminary work mentioned briefly in Ref. \cite{salkin2016generating}
, where the existence of a dripping regime at the lowest air-speeds was mentioned \cite{clanet1999transition}
Using different soap solutions to stabilize our bubbles, we show that the main difference between a pure surfactant and a commercial dishwashing liquid is the bubble lifetime rather than the bubble size.

%
\section{Experimental method} \label{sec:setup}

\begin{figure}[h]
\centering
\includegraphics[width=\linewidth]{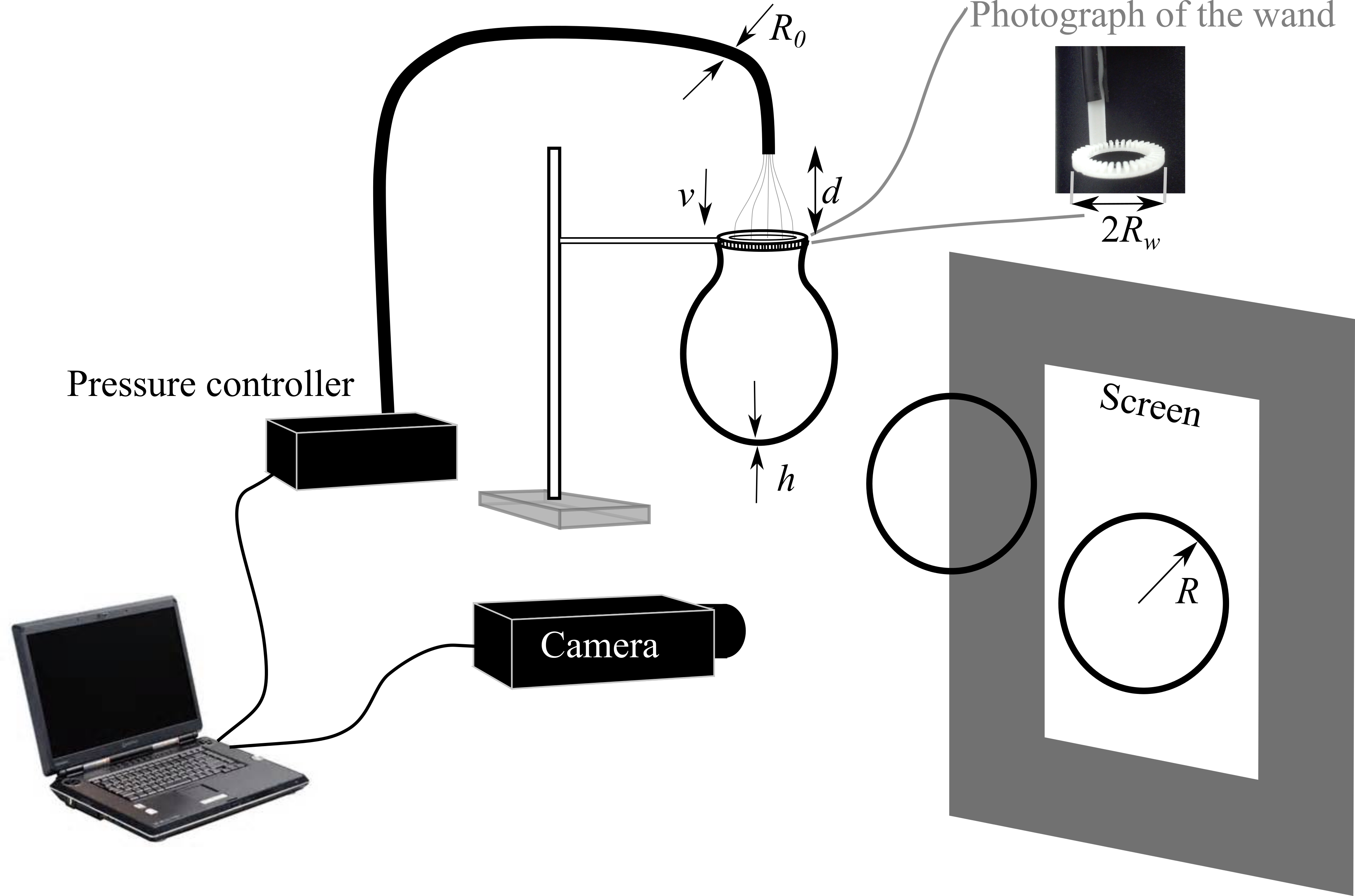} 
\caption{Experimental setup: a bubble is blown by passing air at a controlled pressure through a wand dipped in soap solution. The distance $d$ between the tube orifice and the wand is fixed such that the diameter of the air flow is equal to the diameter of the wand. Images of the bubble are recorded by a digital camera and analysed with ImageJ to extract the radius of the circle of equivalent cross-sectional area.
}
\label{fig:Setup}
\end{figure}

The experiments presented here have been performed using both high-purity research-grade chemicals and a commercially available, multi-component dish-washing liquid. 

The pure system was a sodium dodecyl sulphate (SDS)/Glycerol solution. SDS (Aldrich, France used as received) was dissolved in distilled deionised water (MilliQ conductivity $\sigma = 18.2 $M$\Omega$.cm) to give a range of molar concentrations, generally above the critical micellar concentration $c_{\textrm CMC} \approx$ 8 mM \cite{Zana1993} with most experiments performed with a solution of concentration $c=15$ mM.
As SDS is known to hydrolyse over time producing dodecanol, all samples were used within a week of preparation. 
The SDS powder itself was kept for up to one month before a fresh supply was used.
Glycerol (Sigma Aldrich, France, purity$> 98.5$  \%) was added with a concentration of 20~\% by mass. 
The sample was mixed with a magnetic stirrer until thoroughly homogenised. 
The solution with 15 mM SDS and 20~\% glycerol 
was characterized with commercial apparatus.
The 
surface tension $\gamma = 37.4$ mN/m 
was measured using the pendant drop method (Tracker, Teclis, France)
and the viscosity $\eta = 1.8$ Pa.s 
using a double-couette rheometer (Anton Paar)
.

For the commercial dishwashing product, we used Fairy (also known as Dreft) (Procter and Gamble, Belgium), the exact composition of which is unknown, but it does contain 15-30~\% anionic surfactants (e.g. sodium laureth sulfate, similar to SDS) and 5-15~\% non-ionic surfactants (e.g. lauramine oxide).
We used a concentration of 10 \% by volume.
For these solutions, we measured a surface tension of 25.3 mN/m. The viscosity is the same as water.

The experimental setup is illustrated in Figure \ref{fig:Setup} with a downward airflow directed onto the soap film. The air flow was controlled using an Elveflow OB1 multi-port pressure controller connected to the lab compressed air supply, with pressures between 320 mbar to 480 mbar giving air velocities between 6 m/s and 8 m/s. The air was passed along $2R_0 = $3 mm diameter tubing and exited from an orifice of the same diameter.
At $d=5$ cm downstream from the nozzle, the fully turbulent jet expands to a diameter of 20 mm, where the size of the jet was comparable to $2R_w$ the diameter of the 3D printed wands \cite{salkin2016generating}.
This value of $d$ was then fixed for all subsequent measurements. At this distance, the speed of airflow was measured using a TPI 575 digital hot wire anemometer positioned in the centre of the airflow 
and found to be constant across the surface of the wand
.

Different wands were trialled, including hand-built wire loops of various diameters (using wire of diameter 1 mm) and commercially available plastic wands with ridged edges (which act as a solvent reservoir) before choosing to manufacture custom wands using a 3D printer, with radius $R_w$ of 9.5 mm, width of 4.5 mm, thickness $e$ of 3.3 mm and patterned with 36 periodic ridges fanning out from the inside to the outside. 
Additionally, the material used by the 3D-printer (ABS) is slightly porous.
For these reasons, the effective length of the wand, on which the contact line is pinned, is unknown. We thus introduce a parameter $r$ defined as the ratio between the effective length and the measured wand perimeter.  

Complementary experiments were undertaken using a Vortice VC1 electric blower to generate upwards airflow, controlled using a Griffin and George variable AC transformer on the input voltage. This setup necessitated larger tubing, a nozzle of diameter 7 mm and hand-built wire loops of diameter 15 mm. Air speed was also measured using a hotwire anemometer. The wider nozzle allowed for measurements to characterise the air velocity within the expanding jet 

Images of the bubbles were recorded using digital cameras (either Imaging Source DBK 41AF02 or UEye U148SE) at frame rates of up to 30 frames per second.
Bubbles were either illuminated with ambient laboratory lighting and blown in front of a dark background, or illuminated from behind using a LED light panel (DORR LP-200LED 17.8 $\times$ 12.7 cm). Images were analysed using ImageJ software package, and the bubbles characterised by a radius of equivalent cross-sectional area, as they were not perfectly spherical due to the air flow.

\section{Results} \label{sec:results}

\begin{figure}
\centering
\includegraphics[width=\linewidth]{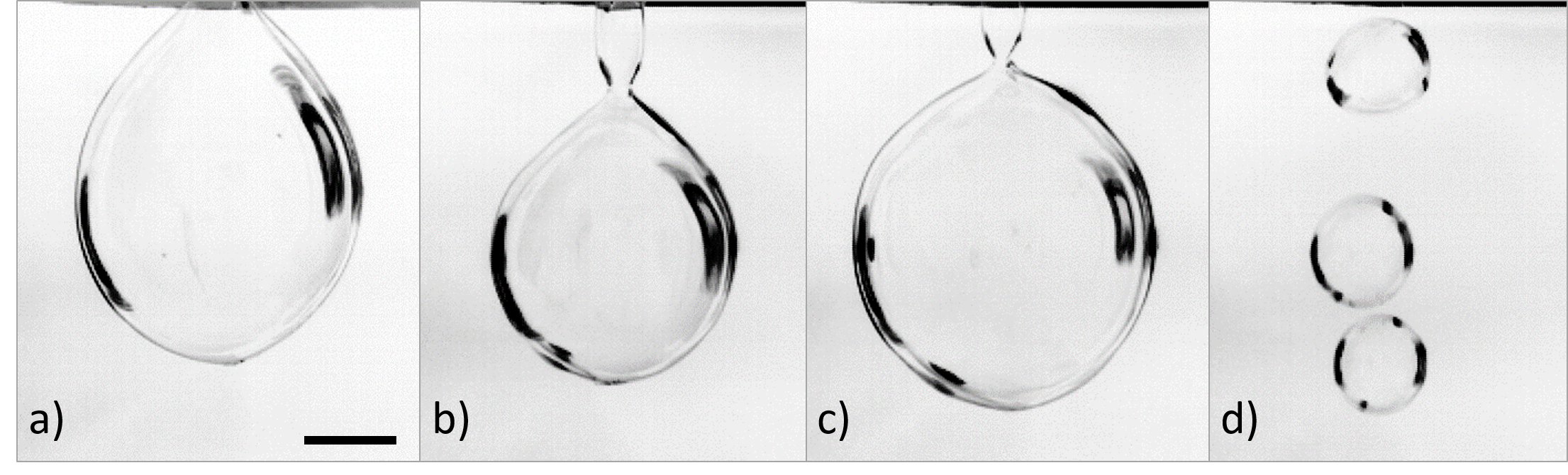} 
\caption{Categories of bubbles 
with increasing air speed: (a) 
v=7.5 m/s,
open bubbles, which burst before detaching; (b) 
v=8.0 m/s,
almost closed bubbles which form a narrow neck but do not quite detach; (c) 
v=7.8 m/s,
closed bubbles which successfully detach from the wand; 
(d) v= 9.2 m/s, small bubbles obtained in the jetting regime. Images have been inverted to emphasise the interface. The scale bar is 5 cm.}
\label{fig:BubbleTypes}
\end{figure}

\begin{figure}
\centering
\includegraphics[width=\linewidth]{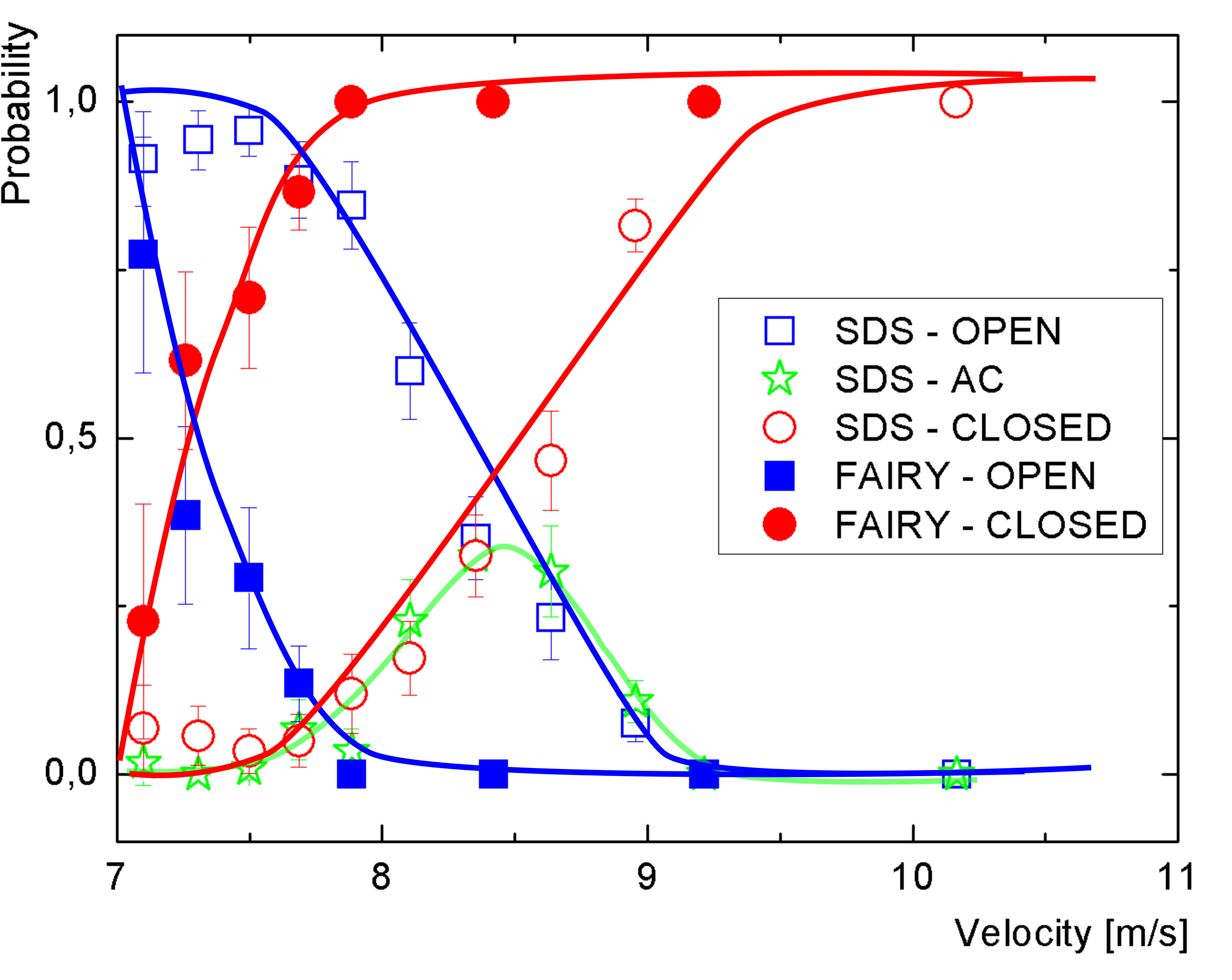}
\caption{The probability of obtaining open, closed and almost closed (AC) bubbles plotted against air velocity, for both 15 mM SDS/20 \% Glycerol and Fairy solutions. Almost closed bubbles are only seen in SDS solutions, in a narrow velocity range. The  lines are guides for the eyes.  }
\label{fig:prob-type}
\end{figure}

As the air speed was increased from zero the soap film became progressively more deformed. No bubbles were observed until a threshold air speed $v_0$ was reached. Beyond this threshold, we observed three different types of bubble, depending on when they burst, depicted in Fig.\ref{fig:BubbleTypes}.
Those which burst before detaching from the wand are described as {\em open bubbles} and those which detach and float away from the wand are {\em closed bubbles}. 
Between these, we identify {\em almost closed bubbles} which very nearly detach, and show a narrow neck, but do not quite separate completely from the wand before bursting. As can be seen in Fig.\ref{fig:BubbleTypes}, the bubbles are significantly larger than the wand diameter, 1.9 cm.
In Fig.\ref{fig:prob-type} we plot the probability of formation of each bubble type at varying air speed for both SDS/Glycerol and Fairy solutions.

With the Fairy solution, we observe no bubbles at air speeds below 7 m/s. There is an abrupt transition from open to closed bubbles with a well defined closed-bubble transition velocity $v_c = 7.3$ m/s, where the probability values for open and closed bubbles are equal. With this solution, we observe hardly any almost closed bubbles. 

For the solutions of SDS/Glycerol, the first bubbles also appear at a threshold velocity of around $v_0=7$ m/s. However, most of these bubbles burst before detaching (open bubbles). On increasing the speed, we see a broad transition region from open to closed bubbles with up to 30 \% almost closed bubbles. The closed-bubble transition velocity is $v_c \approx 8.5$ m/s. 
For solutions at lower concentrations (roughly below the cmc, data not shown) no bubbles of any type are seen. For solutions with higher SDS concentrations we find that $v_c$ decreases with concentration (data not shown).

\subsection{Bubble Size}
\begin{figure}[!ht]
  \centering
    \includegraphics[width=\linewidth]{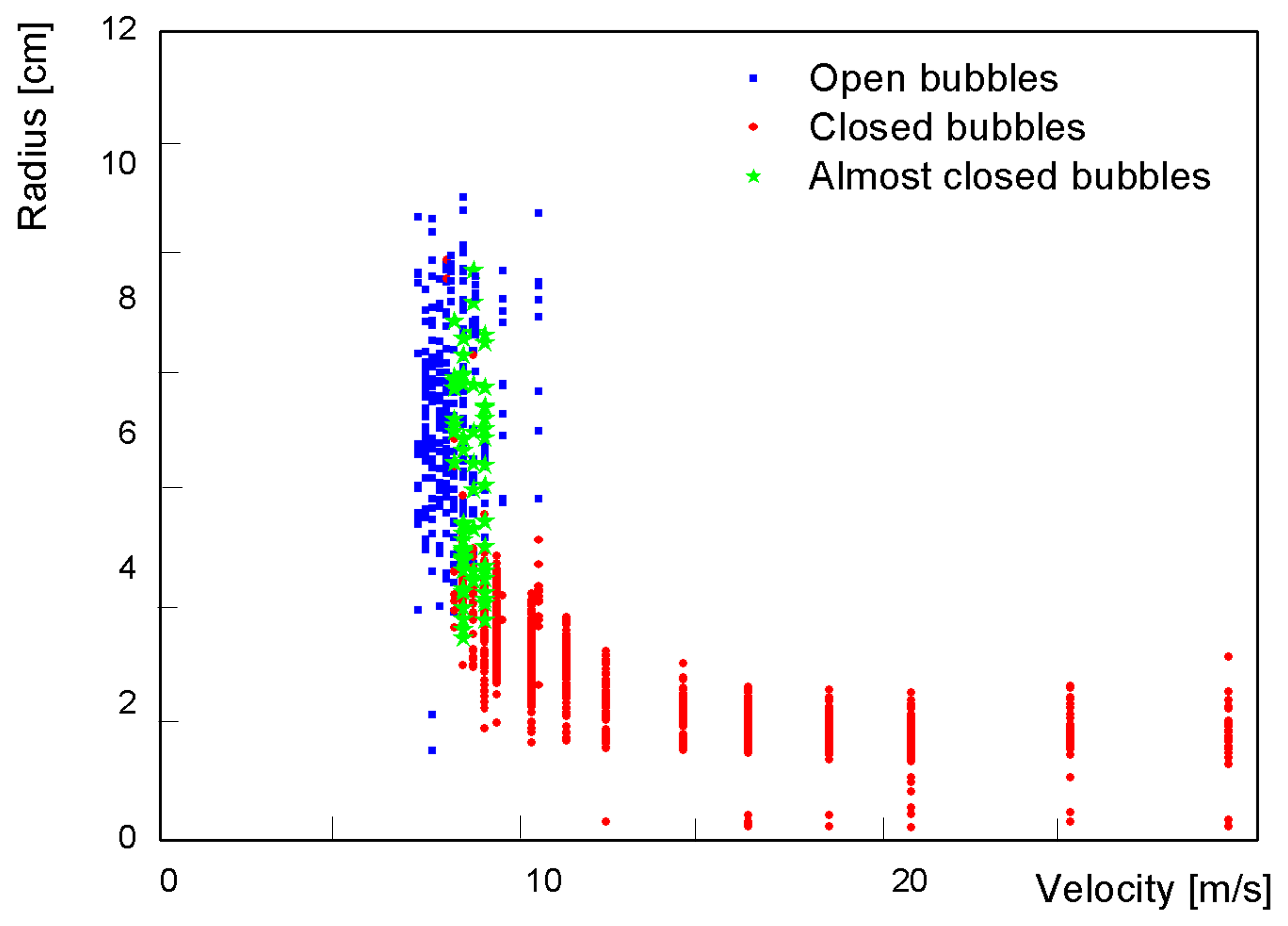}
  \caption{Radius of the bubbles obtained for different air velocities using a
solution of SDS at 15 mM with 20\% glycerol.  The blue symbols represent
the size of the bubbles which burst before detaching (open bubbles) and
the red symbols the size of the bubbles which burst after detaching from the
wand (closed bubbles). The green stars correspond to almost closed bubbles. 
} 
  \label{fig:SDS}
\end{figure}

In addition to characterizing the bubbles as open or closed, we also used ImageJ to measure the equivalent radius of every bubble. Fig.\ref{fig:SDS} shows the sizes of over 1500 bubbles blown using 15 mM SDS with 20\% glycerol. We see significant scatter in the measured bubble sizes, presumably caused by stochastic bubble rupture but also by  fluctuations in air speed and in the thickness of the soap film on the wand. 
Despite the variability, we can make some general observations. 
The largest bubbles, with radius of almost 10 cm, are formed around $v=7$ m/s and are typically open or almost closed. 
On increasing the air speed, the bubbles become smaller with an increasing probability of being closed. 
Around $v\approx$10 m/s there is a clear transition in bubble sizes, which now no longer depend on $v$ and exhibit a radius $R$ around 2 cm. 
This transition corresponds to the the dripping to jetting transition seen by liquids flowing through a narrow orifice \cite{clanet1999transition}, the existence of which Salkin also showed with bubbles \cite{salkin2016generating}. 

\begin{figure}
\centering
\includegraphics[width=\linewidth]{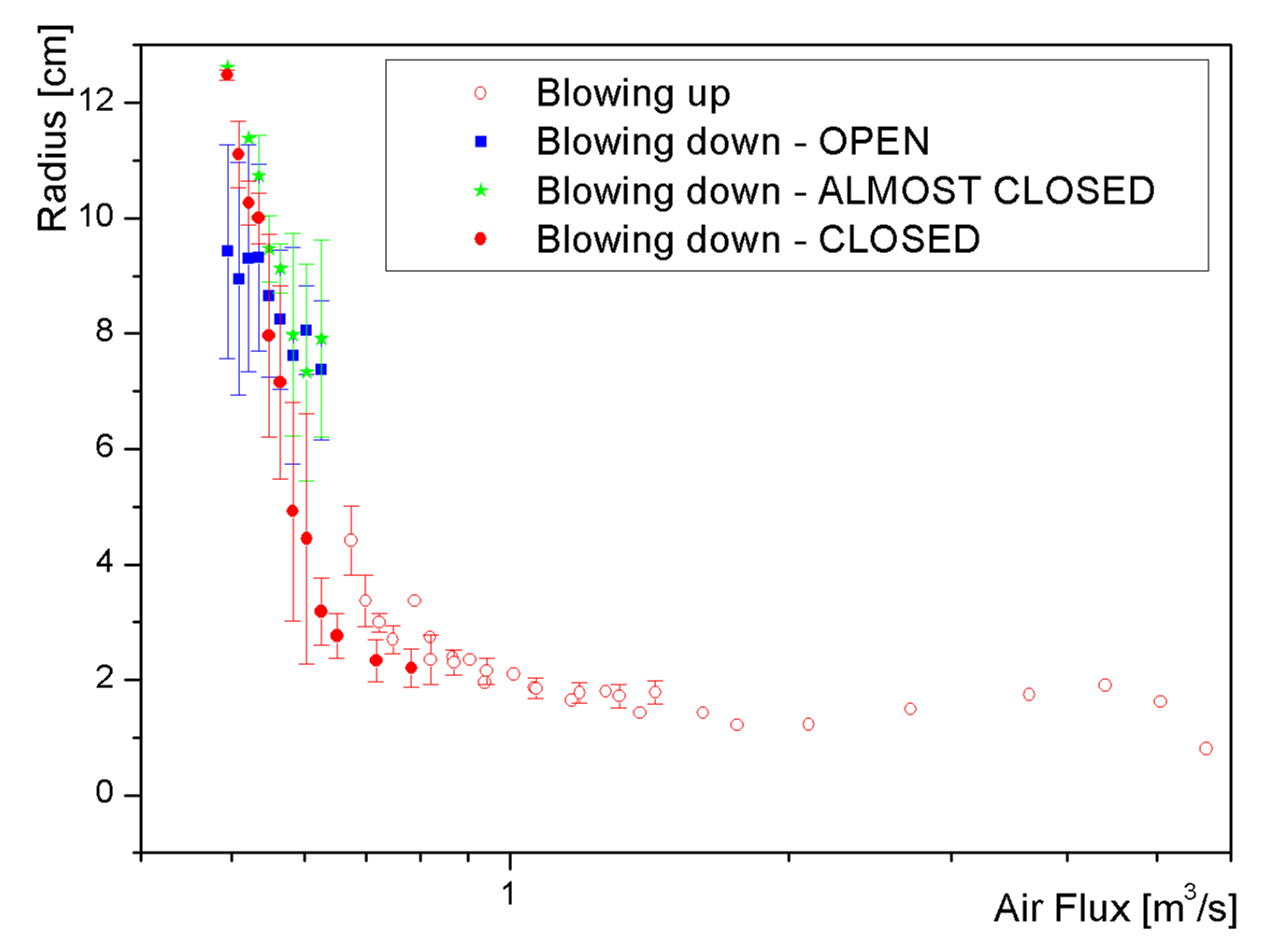} 
\caption{Comparing the bubble size using both up and down air flow - the up direction used larger nozzles, so the wind speed is normalised by calculating the flux (i.e. velocity $\times$ area). 
}
\label{fig:UpDown}
\end{figure}

When comparing these downwards airflow results with those from experiments using a larger nozzle and upwards airflow, we found a systematic difference in threshold wind speeds. By converting wind speed to total air flux (by multiplying by the cross-section area of the respective nozzle) the data collapses, as shown in Figure \ref{fig:UpDown}. The high speed region is now extended to significantly higher fluxes. 

\begin{figure}[!ht]
  \centering
    \includegraphics[width=\linewidth]{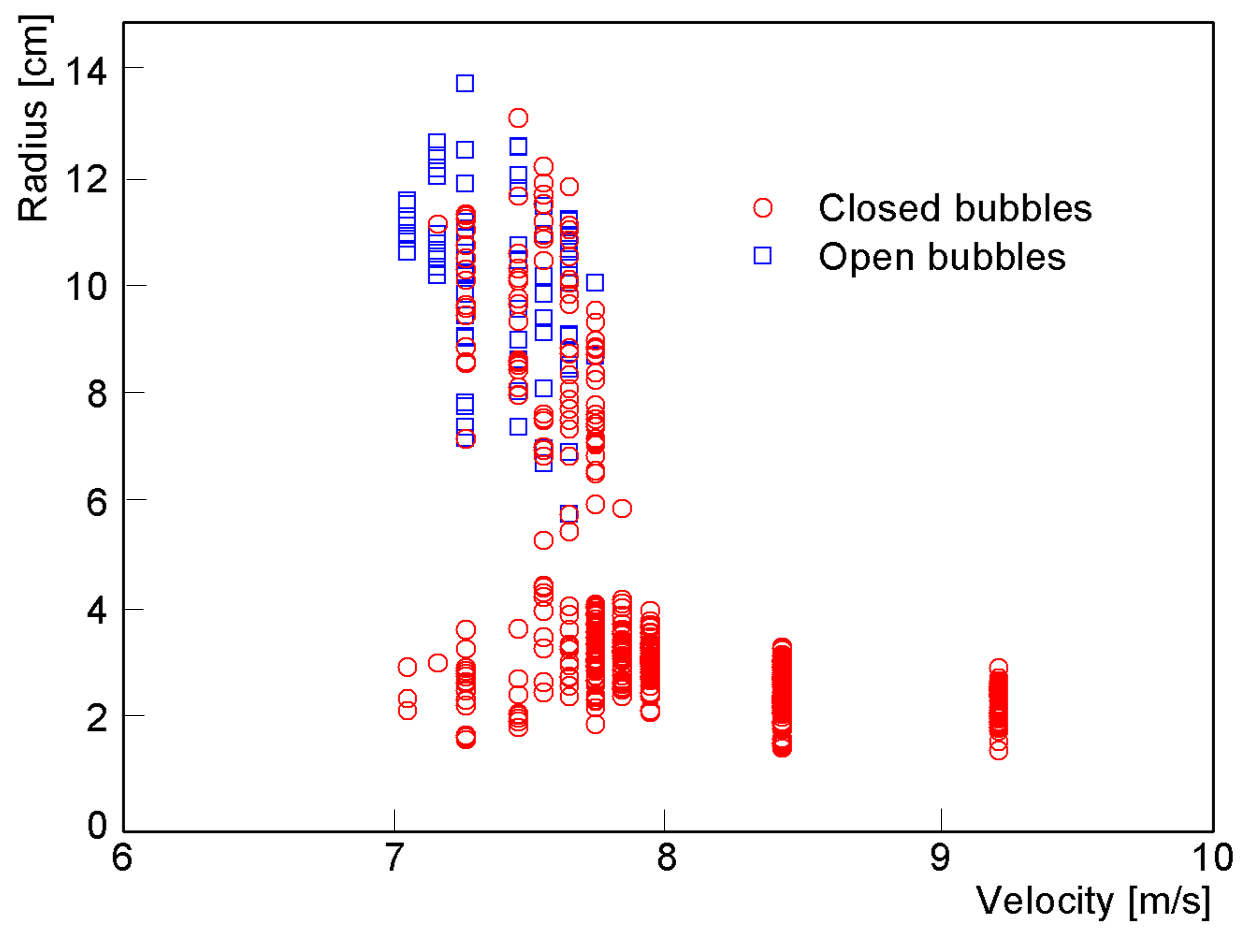}
  \caption{Radius of the bubbles obtained for different air velocities using a
solution of commercial detergent (Fairy).  The blue symbols represent
the size of the bubbles which burst before detaching (open bubbles) and
the red symbols the size of the bubbles which burst after detaching from the
wand (closed bubbles). 
}.
  \label{fig:Fairy}
\end{figure}

For bubbles made with the commercial solution, Fairy liquid, all the measured bubble radii as a function of wind speed are shown in Figure \ref{fig:Fairy}. 
As with the SDS bubbles, two populations of bubbles are seen: (i) at high air velocity, small bubbles with an average radius of around 2 cm whose size remains constant with changing air speed and (ii) at low air velocity large bubbles whose size decreases with air speed. 
Unlike with SDS, the Fairy solutions show that at lower air speeds it is possible to create either small or large bubbles: at a given speed the two populations coexist, but are still distinct.
Additionally, the large bubbles have a much higher probability of being closed compared to those obtained with SDS (see also Fig. \ref{fig:prob-type}).

\section{Discussion} \label{sec:disc}

\subsection{Threshold velocity}

At very low air speeds the film hardly moves, but as the speed increases the kinetic energy of the moving air becomes sufficient to deform the film. 
The threshold air speed $v_0$ to overcome the curvature energy can be calculated from the maximum Laplace pressure required during deformation of the soap film, which is the value predicted by Eq. \ref{eq:V_0Salkin} \cite{salkin2014creation}.

Using relevant experimental values ($\gamma_\text{SDS} = 39$ mN/m, $\gamma_\text{Fairy} = 25$ mN/m, $\rho_g$ = 1.2 kg/m$^3$, $R_w$ = 9.5 mm and $d = 5$ cm) gives $v_0\approx 6$ m/s for SDS/Glycerol and $7.3$ m/s for Fairy, which is a little lower than the experimental value for SDS/Glycerol but corresponds well to the threshold air speed observed in Fig \ref{fig:prob-type} for Fairy, where $v_0 \approx 7$ m/s. One reason for the underestimation of the threshold in the case of SDS/Glycerol may be that the bubbles obtained at such small velocities burst during their formation before they have been measured.

\subsection{Bubble Types}
As the surface tension values of SDS and Fairy are similar, we observe a subsequent similarity in threshold velocities. However surface tension alone is not sufficient to fully understand and predict the bubble-blowing process of a given surfactant. Figure \ref{fig:prob-type} illustrates that there are significant differences between the two solutions we considered. 
For the solutions of Fairy, at air velocities below $7.25$ m/s more than 50 \% of the bubbles are open. 
Between 7.5 and 8.0 m/s, although most of the bubbles are closed, some of them break before detachment, with a probability of breaking around 0.3.
Above a velocity of around 8.0 m/s, the bubbles become almost exclusively closed, with a probability approaching 1. 
At this velocity, the bubbles are rather small and correspond to the jetting regime (Fig. \ref{fig:Fairy}).

This is very different to  what is observed for bubbles stabilized by SDS/Glycerol, for which only a few big bubbles are ever closed.
In particular, a probability close to 1, corresponding to mostly closed bubbles, is reached only in the jetting regime when the bubbles are small.

This emphasises that the big bubbles stabilized by SDS are much more prone to burst than the ones stabilized by Fairy liquid. 
The poorer stability of the SDS bubbles is not surprising (street artists never use SDS in their solutions), however we note that the creation of large bubbles is possible, despite their bursting before detachment. 

A better understanding of this observation is still an  open question (and is known to depend on added high molecular weight polymers \cite{frazier2020make}) and is beyond the scope of this paper in which we chose to concentrate on the description of the bubble size $R$, whether they burst or not before detachment.

\subsection{Dripping to Jetting}
In 1864, Tate was the first to consider the formation and size of droplets dripping, for example, from a tap \cite{tate1864xxx}. His measurements were well described by a simple mathematical model that balanced the droplet mass with the surface tension. Even earlier than this, in 1833, Savart \cite{savart1833veinesliquides} investigated the instability of a falling liquid jet, observing it to break up into small droplets.

At high air speeds, our small bubbles, with a radius around 2 cm, correspond to the jetting regime observed and described by Salkin \textit{et al.} \cite{salkin2016generating}.
The size is due to the Rayleigh Plateau instability \cite{salkin2014creation} and the bubble radius is twice the radius of the wand (horizontal lines in Fig. \ref{fig:SDS} and  \ref{fig:Fairy}). 
In the following, we concentrate on the largest bubbles made at the lower air speeds, 
the existence which was noted by Salkin et al. \cite{salkin2016generating}. We characterise the transition and propose that the large bubbles do not form as a consequence of instabilities in a hollow soap tube, but are inflated while attached to the wand and detach at some criteria.
We propose a mechanism similar to that proposed by Tate for liquid droplets. 

\subsection{Dripping mode}
 
There are three main forces that act during the downward-blowing experiments, in which we observe the dripping mode: an inertial force, the weight of the bubble and surface tension.

The inertial force $I$ is directed downward. It is due to the moving air which expands the bubble at a rate of $2\dot{R}$ directly beneath the wand so we can write $I = \frac{1}{2} \rho_g (2\dot{R})^2 \pi R_w^2$ .

To estimate the inertial force, we follow the method of Clerget et al. \cite{clerget2020bubbleshrink} and write the Bernoulli equation along a stream line between the center of the wand and the expanding edge of the bubble, which gives
\begin{equation}
P_0+\frac{4 \gamma}{R}+\frac{1}{2}\rho_g (2\dot{R})^2 = P_0     + \frac{1}{2} \rho_g   v^2.
\end{equation}
Now the inertial pressure $I$ can be expressed in terms of the the inertia of the jet and the Laplace pressure as 
$I=\left( \frac{1}{2}\rho_g v^2 - \frac{4\gamma}{R}\right) \pi R_w^2$.

The weight of the liquid contained in the film surrounding the bubble also acts downwards and depends on the bubble surface area and the thickness $h$ of the soap film. Assuming a spherical bubble, this can be written as $M = 4 \pi R^2 \rho_l g h $ where $\rho_l$ is the density of the liquid.

The surface tension force acts upwards, keeping the bubble attached to the wand. It acts along the  length over which the film is in contact with the wand, equal to wand's perimeter multiplied by a parameter $r$ which quantifies both the macroscopic ridges and the microscopic porosity of the wand. So, finally, we write the surface tension force as
$T=4 \pi r \gamma R_w$. If we neglect the roughness ($r=1$) we find that the surface tension is never high enough to compensate for inertia.

\begin{figure}[!ht]
  \centering
    \includegraphics[width=\linewidth]{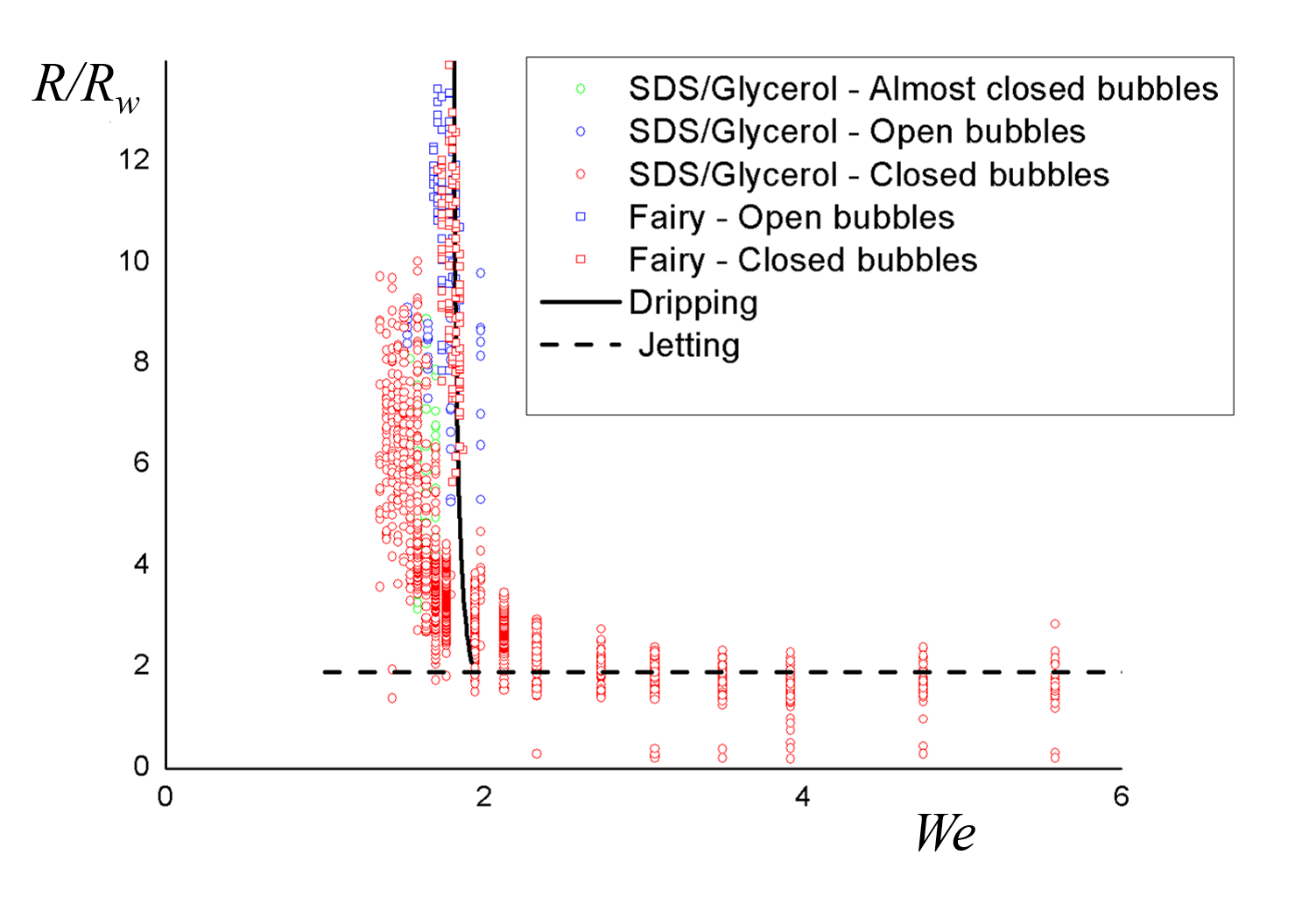}
  \caption{
  Bubble radius normalized by the wand velocity plotted as a function of the Weber number. The dashed line corresponds to a jetting model, in which $R=2 R_w$ and the solid line is the best fit by Equation \ref{eq:dimensionless} of the entire set of data.}
  \label{fig:AllDataDimensionless}
\end{figure}

Both $I$ and $M$ act to detach the bubble, while $T$ keeps the bubble attached to the wand provided $I+M \le T$.
If we estimate these three terms with a film thickness around 1 $\mu$m, the weight is around 8 times smaller than inertia. Additionally, the film thickness, although probably micrometric is unknown. We thus choose to neglect the weight which leads to
\begin{equation}
v=\sqrt{\frac{8\gamma}{\rho_g}\left( \frac{1}{R} + \frac{r}{R_w}\right)}.
\label{eq:velocity}
\end{equation}
Equivalently, we can express this in term of dimensionless variables by introducing the Weber number $We = \frac{\rho_g v^2 R_w}{8 \gamma}$ and by normalizing the bubble radius by the wand radius. This leads to the simple expression of 
\begin{equation}
   \frac{R}{R_w} = \frac{1}{We-r}  \label{eq:dimensionless}
\end{equation}
This prediction is plotted using $r$ as an adjustable parameter together with data 
in Figure \ref{fig:AllDataDimensionless}.
The data obtained with the Fairy and with the glycerol collapse on a master curve and their size
is well described by this model using a value of $r$ equal to 3.2 for both SDS and Fairy/Glycerol solutions, showing that our model captures the main physical parameters.

Note that the agreement between data and model is very sensitive to the value of $r$. 
This could explain why people use a rough wand to blow bubbles: as well as serving as a reservoir for the solution, it provides additional contact area to hold the bubble in place.
The agreement is also very sensitive to the surface tension, which is not surprising since the pendant drop experiment uses the dripping mode of droplets to measure this quantity. An alternative interpretation of $r$ could thus be a larger surface tension, which would give a slightly different equation since the surface tension acts in both terms of Eq. \ref{eq:velocity}.
Both a different surface tension and an effective wand length could also contribute in principle.

We have shown that gravity is negligible in predicting the maximum size of the bubbles in our dripping model.
Thus we could expect that big bubbles can be formed by blowing up, which is not the case as shown in Figure \ref{fig:UpDown}.
We propose that gravity driven drainage cannot be neglected in this case. 
When blowing up, the film at the top of the bubble is expected to thin and eventually bursts.
On the contrary, when blowing down, the film at the top is fed by the liquid contained in the wand.

Interestingly Zhou \textit{et al.} \cite{zhou2017formation}, using an experimental setup similar to Salkin's, have observed the opposite of our findings: smaller bubbles at lower air speeds, larger at higher speeds and a regime between in which no bubbles are observed. However, their air speeds (as characterised by the Reynolds number) are slower than ours, meaning that their no-bubble regime may correspond with our almost closed bubbles regime.

\section{Conclusion} \label{sec:conclusion}
%

To conclude, we measured the size of bubbles generated by controlled blowing on a film of soap solution suspended in a rough wand.
The measured threshold velocity necessary to blow a bubble agrees with previous predictions as do the bubble sizes observed at high velocities, in the Rayleigh-Plateau regime, which are the same size as the wand diameter.
At lower velocities, when blowing downwards, the bubbles are formed one by one in a dripping mode. We create bubbles that are significantly larger than the wand by blowing at low velocity, very near the threshold.
In this regime, the bubble size is well described by a balance between inertia and surface tension.

We observed different types of bubbles: open, closed and almost closed. 
The latter are only observed for SDS/Glycerol solutions. 
Nevertheless, the bubble size seems independent of the bubble type so that it is possible to blow big bubbles with SDS solutions but we never observe their detachment.

In practice, big bubbles are actually blown by varying the air speed - high initial to overcome Laplace pressure, then gently inflating a bubble without detaching at low speeds, and then increasing sharply to detach. This observation provides scope for future work.

\bibliographystyle{plain}
\bibliography{bubbly}

\end{document}